\documentclass{appolb}
\usepackage{slashed}
\usepackage{graphicx}
\usepackage{color}

\begin{document}
\title{Compact Stars, Heavy Ion Collisions, and \\ Possible Lessons For QCD at Finite Densities %
\thanks{Presented at the 'HIC for FAIR Workshop and
XXVIII Max Born Symposium Satellite meeting for QM 2011' in Wroc{\l}aw, Poland.
 }%
}
\author{Thomas Kl\"ahn$^1$, David Blaschke$^{1,2}$, R. {\L}astowiecki$^1$
\address{$^1$Institute for Theoretical Physics, University of Wroc{\l}aw, Poland\\
$^2$Bogoliubov Laboratory for Theoretical Physics, JINR Dubna, 
Russia}
}
\maketitle
\begin{abstract}
Large neutron star masses as the recently measured $1.97\pm0.04$ M$_\odot$ for
PSR J1614-2230  provide a valuable lower limit on the stiffness of
the equation of state of dense, nuclear and quark matter. 
Complementary, the analysis of the elliptic flow in heavy ion collisions suggests an 
upper limit on the EoS stiffness.
We illustrate how this dichotomy permits to constrain
parameters of effective EoS models which otherwise could not be derived 
unambiguously from first principles.
\end{abstract}

%
%
\section{Introduction}
Neutron stars (NS) which outrange the domain of 'typical' NS masses (very 
roughly between 1.2 and 1.6 M$_\odot$, see Fig.~1 in \cite{Lattimer:2010uk} 
for an overview)
have been considered to provide a serious constraint on the stiffness of the 
equation of state (EoS) \cite{Klahn:2006ir}
but treated rather cautiously due to the fact that in the rare cases of very 
massive observed NS
either the accuracy of the measurement has been worryingly low or/and the 
measurement itself  raised doubts within the community \cite{Lattimer:2006xb}.
It seems, that this situation has changed after a mass of 
M=$1.97\pm0.04$ M$_\odot$ 
has been reported for PSR J1614-2230 with an unprecedented accuracy in this 
high mass regime \cite{Demorest:2010bx}
and without perceivable objections from expert groups involved in this field.
This observation of a two solar mass NS provides a very direct constraint on 
the minimum stiffness of the EoS
of cold and dense matter and therefore promises new insights regarding our 
understanding in particular
of {\it the nature of particle interactions at finite densities}.
Implications have been discussed shortly after the measurement became public 
\cite{Lattimer:2010uk}.

As an example (which deserves more attention than can be granted in this 
article) for the importance of this result
it is worth to mention the problem of strangeness for the hadronic EoS.
Even though we are rather interested in the possible existence of a quark 
matter (QM) core in NS one can not exclude the possibility
that before the according critical density can be reached the mass-density 
threshold for hyperons is passed
and the EoS has to soften due to the appearance of these additional degrees of 
freedom.
Due to this softening the maximum mass of a corresponding NS 
can decrease drastically in comparison to the plain EoS with only neutron and 
proton degrees
of freedom and is therefore not unlikely to contradict the high NS mass 
constraint.
The problem can be avoided by accounting for repulsive vector interaction
terms which are a specific feature of relativistic approaches to EoS.
This has been shown, e.g., in a generalized nonlinear Walecka-type model 
\cite{Bednarek:2009at,Bejger:2011bq}
and a quark-meson-coupling model \cite{RikovskaStone:2006ta,Stone:2010jt}.
Similarly, it turns out that vector interaction terms are of crucial importance
if aiming at the description of high mass NS with QM cores \cite{Klahn:2006iw}.
It is an interesting finding that merely by the observation of a two solar 
mass NS
the vector interaction can be identified as an inevitable channel if 
approaching a microscopical
description of both, dense hyperon and QM.

Besides a two solar mass constraint limiting the 'softness' of the
EoS we suggested several other constraints from NS observations
and the analysis of heavy ion collisions which have to be fulfilled 
{\it simultaneously}
by a viable state-of-the-art EoS \cite{Klahn:2006ir}.
In detail, the scheme suggests that a viable EoS should
\begin{itemize}
\item reproduce the most massive observed neutron star,
\item avoid the direct URCA (DU) cooling problem,
\item result in neutron stars within the predicted mass-radius domains
  of 4U 0614+09 (deduced from quasiperiodic brightness
  oscillations) and RX J1856-3754 (deduced from the objects thermal emission),
\item explain the gravitational mass and total baryon number of
pulsar PSR J0737-3039(B) with at most 1\% deviation from the baryon
number predicted for this particular object, and
\item not contradict flow and kaon production data of heavy-ion
collisions.
\end{itemize}
At the time of publication of Ref.\ \cite{Klahn:2006ir}, the most massive 
NS has been  PSR J0751+1807 with 
$M \sim 2.1$~M$_\odot$, a result which later has  been withdrawn 
\cite{Nice:2007}. 
We reasoned, that in particular the flow constraint is
an extremely useful constraint.
In contrast to the two solar mass constraint it limits 
the EoS such, that it cannot exceed the upper limits
on the pressure (as a function of density) in symmetric
matter as obtained from the analysis of the elliptic
flow of iso-symmetric matter in HIC \cite{Danielewicz:2002pu}.
This gives two complementing phenomenological constraints
on the EoS stiffness, an upper (flow) and a lower (two solar mass constraint) limit, 
which both considered together significantly reduce the possible shape of
the pressure density relation, viz., the EoS at super saturation densities.
In previous work we took advantage of this insight
and adjusted otherwise not well determined coupling constants of a 
Nambu--Jona-Lasinio (NJL) type model.
Details about this models framework is found in the original work formulated 
before the invention of the quark model of elementary particles
\cite{Nambu:1961tp,Nambu:1961fr}
and a number of review articles 
\cite{238312,Vogl:1991qt,343506,hep-ph/9401310,hep-ph/0402234}
applying it to elucidate the role of dynamical chiral symmetry breaking 
for hadron structure and its restoration in dense quark matter.
Our resulting hybrid,
nuclear-quark matter EoS, is in agreement with both
of these before mentioned constraints and shows a better
overall performance than the originally underlying nuclear EoS 
based on the Dirac-Bruckner Hartree-Fock (DBHF) approach \cite{Klahn:2006iw},
which will be shortly summarized later in this paper.
At this time we had to scan a small parameter range only
in order to obtain this result.
However, the question how well the QM model EoS
is constrained, viz. which other sets of coupling constants would
reproduce a phenomenologically sound EoS, has been left open in 
\cite{Klahn:2006iw}.
The reporting of PSR J1614's high mass has triggered
this kind of more systematic studies for Bag-like QM EoS 
within a purely phenomenological model for the QM equation of state
consisting simply of a power series expansion in the quark chemical potential 
$\mu$ including 4$^{th}$ and 2$^{nd}$ order terms 
\cite{Ozel:2010bz,Weissenborn:2011qu}.
The 4$^{th}$-order term is thought to mimic the influence of strong 
interactions
on the ideal gas expression for the quark pressure, while the 2$^{nd}$ order 
term is a measure for the competing effects of a finite strange quark mass and 
a possible diquark condensate in deconfined matter \cite{Alford:2004pf}.
It is worth to notice that a $\mu^4$-term is not necessarily the leading order 
term in a $\mu$-expansion of the pressure, as has been shown for a simple,
semi-analytic model based on Dyson-Schwinger techniques at finite densities 
\cite{Klahn:2009mb}.
We would like to mention a very promising, recent development in 
modeling quark matter in the nonperturbative, low-energy domain of QCD which is
relevant for the QCD phase diagram and compact star phenomenology. 
This concerns nonlocal, separable interaction models with either covariant or
instantaneous formfactors  
\cite{Blaschke:2004cc,arXiv:1012.0664,arXiv:1012.2113} and their 
application to compact stars \cite{Blaschke:2007ri,Grunfeld:2007jt}
generalizing the local current-current coupling of the NJL model.
In particular the rank-2 separable models which allow a simultaneous 
description of the dynamical quark mass function $m(p)$ and the wave function 
renormalization $Z(p)$ of the quark propagator in accordance with lattice QCD 
data \cite{arXiv:0806.0818,arXiv:0810.1099,arXiv:1104.0572,arXiv:1006.4639} 
shall allow to greatly reduce ambiguities of 
the parametrization of quark models discussed in these Proceedings.
For the time being, however, the nonlocal models need to be further improved
to address diquark condensation and asymmetric matter before they can be used
to study compact star constraints.

In this paper we will illustrate in a systematic, 'whole-range' scan of an NJL 
model EoS for QM how strongly a two-solar NS mass measurement constrains the 
strength of available (vector and diquark) coupling strength parameters
and what implications can be derived for the investigation of HIC, viz. the EoS
of iso-spin symmetric matter.
Section \ref{SEC:NJL} gives a summary of the applied QM EoS,
section \ref{SEC:RESULTS} discusses the result of our analysis,
section \ref{SEC:CONCL} will provide conclusions and a brief outlook concerning
further interesting questions.

%
%

%
%

\section{Hybrid Matter EoS based on a NJL Model}
\label{SEC:NJL}

In order to obtain a QM equation of state we employ a three-flavor 
color superconducting NJL model with selfconsistently determined quark masses
and diquark gaps \cite{Ruester:2005jc,Blaschke:2005uj,Abuki:2005ms}.
In addition to a typical scalar interaction term, 
we account for a repulsive vector interaction term
which stiffens the EoS with increasing interaction strength and results in 
sufficiently high NS masses (details are found in \cite{Klahn:2006iw} and 
references therein).
The attractive scalar diquark channels are responsible for the formation of 
diquark condensates and color superconducting phases in the system.
As we discussed before  it moreover lowers the transition density to a nuclear matter EoS
with increasing coupling strength \cite{Klahn:2006iw}.
The effective Lagrangian can be split into a free particle and an interaction part.
The free part reads as
\begin{equation}
{\cal L}_{kin} = \bar{q}(-i\gamma_\mu \partial_\mu + \hat{m} + \hat{\mu})q,
\end{equation}
where $\hat{m}={\rm diag}(m_u,m_d,m_s)$ is the diagonal current quark mass 
matrix and $\hat{\mu}$ the corresponding quark chemical potential matrix.
The effective interaction is written as
\small 
\begin{equation}
 {\cal{L}}_{int} = G_S\eta_D\sum_{a,b=2,5,7}(\bar{q}i\gamma_5 \tau_a \lambda_b C \bar{q}^T)(q^TCi\gamma_5\tau_a\lambda_a q) +  
		   G_S\sum_{a=0}^{8}\left[(\bar{q}\tau_a q)^2 + \eta_V(\bar{q}i\gamma_0q)^2\right] .
\end{equation}
$\tau_a$ and $\lambda_a$ are Gell-Mann matrices in flavor and color space 
respectively and $C$ the charge conjugation matrix.
Here we have omitted interaction channels which do not contribute to the 
thermodynamics at meanfield level, like the pseudosclar isovector channel which
would be required to make the chiral symmetry of this interaction model 
manifest.
The parameter $G_S$ defines the scalar coupling strength 
and can be determined from meson properties in the vacuum.
For this study we apply the parameters obtained for a sharp cut-off 
regularization scheme
from Table III of Ref.~\cite{Grigorian:2006qe}
labeled with '$\infty$' in front of the corresponding table row
\footnote{This parametrization scheme has been implemented in an online tool 
developed by F.~Sandin which also corrects for a mistake in the kaon mass 
formula employed in ~\cite{Grigorian:2006qe}, see 
http://3fcs.pendicular.net/psolver}.
The parameters $\eta_D$($\eta_V$) are defined as the ratio of 
the diquark(vector) and scalar coupling.
Since these two parameters are not fixed by vacuum properties of mesons 
or hadrons we treat them as free model parameters.
In order to investigate thermodynamical properties of the system 
we use the partition function in path integral representation,
\begin{equation}
 Z(T,\hat{\mu}) = \int {\cal D}\bar{q}{\cal D}q\exp\left\{\int_0^\beta d\tau\int d^3x[\bar{q}(i\slashed{\partial}-\hat{m}+\hat{\mu}\gamma^0)q + {\cal L}_{int} ]\right\}.
\end{equation}
Bosonic meson field degrees of freedom can easily be introduced by applying 
corresponding
Hubbard-Stratonovich transformations in all interaction channels.
For the sake of simplicity all of them are treated on the meanfield level,
viz., mesonic fluctuations and higher correlations are neglected.
Minimizing the thermodynamical potential $\Omega=-T\ln Z$ with respect to the 
meanfields then defines the pressure $p=-\Omega$ of the equilibrated system.
As a result of the minimization procedure one obtains a set of coupled gap 
equations which has to be solved selfconsistently.
Finally, the thermodynamical potential reads as
\begin{eqnarray}
\label{EQ:GCPot}
 \Omega(T,\mu) &=& \frac{\phi_u^2 + \phi_d^2 + \phi^2_s}{8 G_S} 
- \frac{\omega^2_u + \omega^2_d + \omega^2_s}{8 G_V}
+ \frac{\Delta^2_{ud} + \Delta^2_{us} + \Delta^2_{ds}}{4G_D}\nonumber\\
 &&- \int{\frac{d^3p}{(2\pi)^3}}\sum_{n=1}^{18}
\left[E_n + 2Tln\left(1 + e^{-E_n/T}\right)\right] + \Omega_{l} - \Omega_0. 
\end{eqnarray}
$\Omega_l$ denotes the contribution of electrons and muons to the 
thermodynamic potential, $\Omega_0$ is the contribution to be
subtracted in order to obtain zero vacuum pressure.
The $\phi_f$ (with $f=u,d,s$) are the chiral condensates corresponding to the three quark flavors
which are obtained from the previously described minimization of the scalar meanfield terms.
Accordingly, on has to account for the vector meanfields $\omega_f$
and pairing gaps $\Delta_{ff^\prime}$.
Under neutron star conditions one additionally imposes 
electric charge neutrality and $\beta$-equilibrium conditions. 
In this paper we disregard the Kobayashi-Maskawa-'t Hooft term in the 
interaction Lagrangian, see Ref.~\cite{Blaschke:2005uj} for a motivation.
Further references and possible consequences arising from the inclusion of 
this term are discussed in \cite{Blaschke:2010vj,Blaschke:2010vd}.

With Eq.~(\ref{EQ:GCPot}) we have a QM EoS available which is still, by 
construction, phenomenological but allows to interpret the influence of 
certain interaction channels on the EoS more specifically than 
generalized power expansions of ideal gas expressions.
The next step in order to describe NS phenomenology is to account for confined 
nuclear matter at lower densities.
Without a unified in-medium-approach for the description of 
nuclear matter in terms of QM degrees of freedom at hand
we, as everybody else, fall back to a two-phase description, 
joining independently obtained nuclear and QM EoS
by performing a phase transition construction.
As common as this procedure is we feel a few comments 
have to be made 
about it
in order to avoid a misleading interpretation of this work.
Our understanding of the nuclear EoS at supersaturation densities is by no 
means more profound than our knowledge about dense QM.
The set of constraints we apply in order to pin down the QM EoS we originally
bundled up in order to constrain the nuclear EoS.
There are no profound additional insights into the 
shape of the 'true' nuclear matter EoS 
since the publication of \cite{Klahn:2006ir} even though
progress has been made in order to gain deeper insights into the physics of 
finite density nuclear systems, e.g., within the framework of 
chiral effective field theory \cite{Lacour:2009ej} and directly from QCD via
lattice simulations \cite{arXiv:1101.1463}. 
While these studies are progressing and may become applicable to high density
nuclear systems in compact stars and heavy-ion collisions in future, we discuss
these systems for the time being from the point of view of parametric 
approaches to the high density EoS. 
Such cold dense EoS studies may be guided by observations of compact stars, 
see \cite{Klahn:2006ir,Ozel:2010fw,Steiner:2010fz}.
From this parametric point of view
the variation of model parameters of a QM EoS in a strict sense
would always require to perform a similar variation of an independently
obtained NM EoS.
All results of this (or any similar) analysis, are likely to change
significantly if the NM EoS is exchanged by a  model with significantly 
different high density behavior.
As an example, softening the NM EoS to a degree where the model
is not any longer in agreement with the two-solar mass constraint
requires to soften the QM EoS as well in order to obtain
a thermodynamically sound phase transition.
This makes it hard, if not impossible, to obtain a hybrid EoS
which would describe a two solar mass NS.
Still the QM EoS by itself is not necessarily in contradiction
with the existence of high neutron star masses.

For our analysis we avoid the problem of a nuclear parameter scan
by applying the Dirac-Brueckner Hartree-Fock (DBHF) EoS 
which has proven to perform reasonably well
for describing nuclear matter saturation properties and kaon data  
\cite{Fuchs:2003zn} as well as NS properties \cite{Klahn:2006ir}
even  though it tends to behave too stiff
above densities of about 3.5 times saturation density.
On the other side, this stiffness occurs in
a region where QM degrees of freedom are not
unlikely to be the only ones which are relevant.
Amongst other reasons we prefer the DBHF EoS 
because it is based on a relativistic and microscopical 
description of many-particle interactions.
It starts from a given free
nucleon-nucleon interaction (the relativistic Bonn A potential) 
fitted to nucleon-nucleon scattering data and deuteron properties. 
In ab initio calculations based on many-body techniques one then
derives the nuclear energy functional from first principles,
i.e., treating short-range and many-body correlations explicitly.
In the relativistic DBHF approach
the nucleon inside the medium is dressed by the
self-energy based on a T-matrix. The in-medium T-matrix
as obtained from the Bethe-Salpeter equation
plays the role of an effective two-body interaction
which contains all short-range and many-body correlations
in the ladder approximation.
As we have shown in the context of hybrid EoS the rather stiff behavior at 
high densities
is not necessarily relevant if the phase transition to QM occurs at
low enough densities of about three to four times saturation density 
\cite{Klahn:2006iw}.

%
%

%
%
\section{Results}
\label{SEC:RESULTS}
\subsection{Hybrid Neutron Stars}

Unlike to previous work, where we applied both, the flow and two-solar mass constraint,
simultaneously in order to obtain a phenomenologically sound hybrid EoS \cite{Klahn:2006iw}
we start this analysis from calculating neutron star configurations
for a wide range of vector and diquark couplings ($\eta_V\in[0.0,0.7]$, $\eta_D\in[0.8,1.15]$).
To further extend the previous study we apply a 
different parameterization for the scalar coupling strength, as well, 
applying the parameters from the row labeled '$\infty$' 
in Table I of Ref.\cite{Grigorian:2006qe}.
As we will show, this choice significantly affects the outcome of the present study.

Before we discuss the overall result of a full variation of 
the two free parameters $\eta_V$ and $\eta_D$ 
we keep one of them fixed at a reasonable value and vary the other.
\begin{figure}[h]
 \includegraphics[scale=0.45,angle=-90]{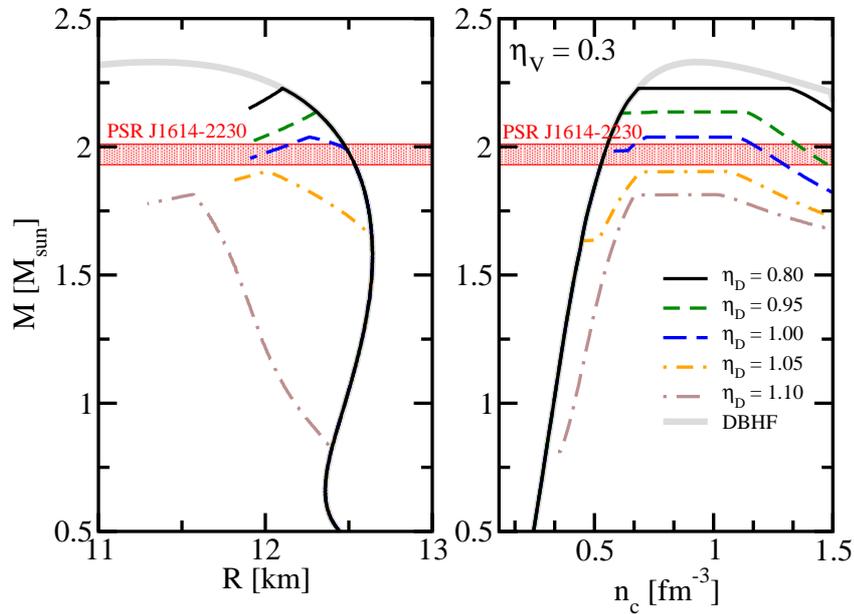}
 \caption{At constant vector coupling (here $\eta_V=0.3$) an increase of $\eta_D$ lowers both, the critical density
and the maximum NS mass.}
 \label{FIG:etaVconst}
\end{figure}
We consider any choice to be reasonable which eventually describes
a two solar mass NS if one varies the one remaining free parameter.
In Fig. \ref{FIG:etaVconst} the vector coupling 
is kept constant at a value of  $\eta_V=0.3$
while the diquark coupling is varied in the range $\eta_D=0.80\dots 1.10$.
At 'low' values of $\eta_D$, 
here up to  $\eta_D=1.00$ we find the required massive NS configurations.
As one observes, an increase of $\eta_D$ does not only result in 
a decrease of the maximum NS mass
but lowers the critical density for the phase transition, too.
In other words, an increase of $\eta_D$ increases the content of QM 
in massive NS and lowers the maximum NS mass at the same time.
\begin{figure}[h]
 \includegraphics[scale=0.45,angle=-90]{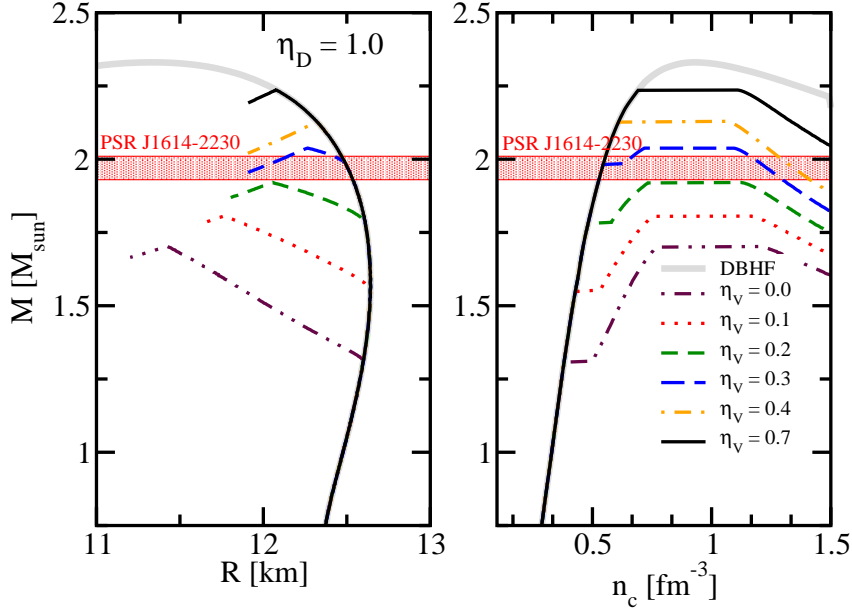}
 \caption{At constant diquark coupling (here, $\eta_D=1.0$) an increase of $\eta_V$ increases the maximum NS mass
and the critical density.}
 \label{FIG:etaDconst}
\end{figure}
On the other hand, keeping the  diquark coupling $\eta_D$ at a constant value 
and increasing the vector coupling will increase both, the maximum NS mass and 
the critical density.
This is illustrated in  Fig.~\ref{FIG:etaDconst}. 
Again, there is a critical value of $\eta_V$ corresponding
to a minimal stiffness of the EoS which has to be exceeded
in order to obtain NS configurations with a sufficiently high
maximum mass. 
In the illustrated example for $\eta_D=1.0$ this holds for values 
larger than a critical coupling slightly above $\eta_D=0.2$.

With this understanding of the influence of $\eta_D$ and $\eta_V$ on
both, the maximum NS mass and the critical density
for the onset of the phase transition we now
perform a variation over the full available parameter space region of $\eta_D$-$\eta_V$
in which we can obtain stable hybrid NS configurations.
\begin{figure}[h]
 \includegraphics[scale=0.45,angle=-90]{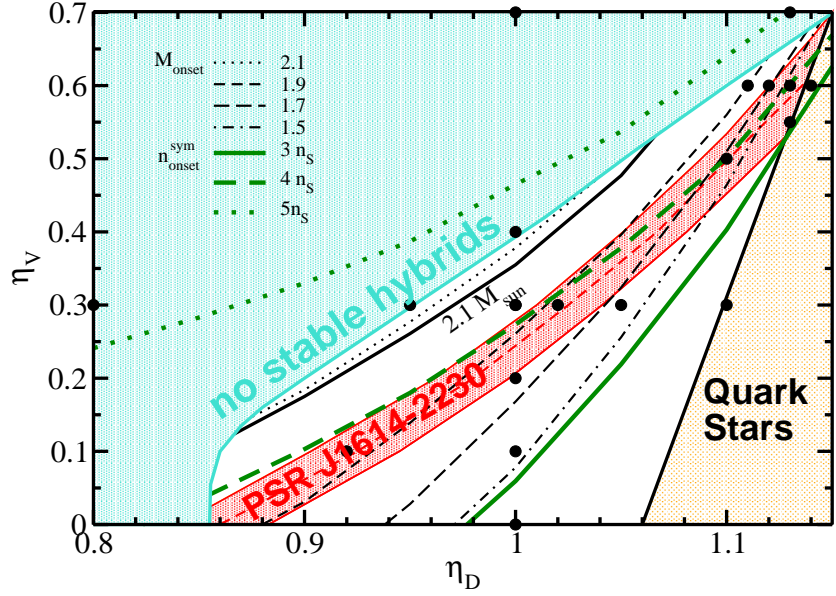}
 \caption{The overall result of the parameter study of the NJL model
concerning the vector ($\eta_V$) and diquark ($\eta_D$) channel coupling.
The red band and all parameter pairs over it correspond
to EoS which reproduce at least $1.97\pm0.04$ M$_\odot$ (PSR J1614-2230),
in the cyan region no stable hybrid configurations are found,
while the grey hatched region corresponds to 'mostly' quark stars.
A more detailed discussion is given in the text. Black filled circles
correspond to parameterizations mentioned in the text.}
 \label{FIG:MonsterPlot}
\end{figure}
The result of this study is summarized in Fig.~\ref{FIG:MonsterPlot},
which we will now discuss in detail.
From the obtained mass-radius and mass-density relations
of each of the differently parameterized hybrid EoS
we extracted two numbers only,
the maximum possible NS mass and the NS mass
at which the central density becomes large enough
to generate a QM core in the NS.
Consequently, we call the latter quantity M$_{\rm onset}$.
The red band in Fig.~\ref{FIG:MonsterPlot} labeled PSR J1614-2230
corresponds to EoS parameterizations
which describe a maximum mass exactly within the
interval M=$1.97\pm0.04$ M$_\odot$ as it has been reported
for PSR J1614-2230.
As we understand from the previous paragraphs
it is possible to obtain more massive solutions
by increasing $\eta_V$ or decreasing $\eta_D$.
As both of these operations increase the transition
density one eventually obtains massive NS
which are purely hadronic. This can be
either because the transition occurs at very
large densities which are not realized in NS
or, more relevant for our considerations,
because the hybrid NS solutions become unstable.
Examples for this situation are found in Fig.~\ref{FIG:etaDconst}
for all $\eta_V>0.4$.
The cyan hatched region in Fig.~\ref{FIG:MonsterPlot}
corresponds to all EoS parameterizations which result in
unstable hybrid solutions.
The opposite extreme scenario results from 
increasing $\eta_D$ or decreasing $\eta_V$.
Then, the transition density is lowered
until eventually only a thin hadronic
layer remains and the NS are basically pure
quark star configurations.
This scenario corresponds to the grey hatched
region labeled 'Quark Stars'.
In this parameter region the quark matter EoS has an early 
onset of the pressure due to a decrease of the dynamical quark mass before 
the first order phase transition which results
in a low QM transition density.

\begin{figure}[h]
 \includegraphics[scale=0.45,angle=-90]{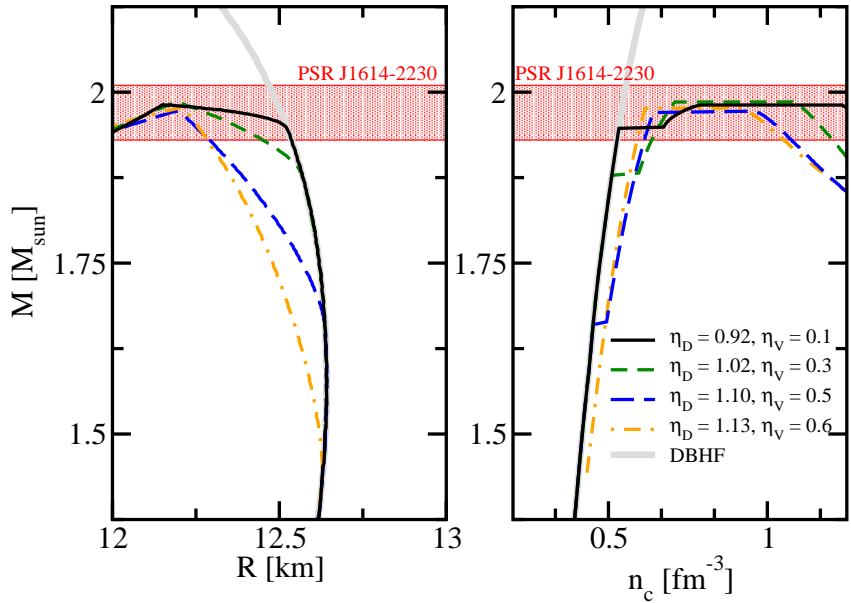}
 \caption{Systematics of mass-radius and mass-central density curves for 
chosen parameter sets with maximum mass equal to that of PSR J1614-2230. 
Large values of $\eta_D$ and $\eta_V$ result in a phase transition at lower 
densities and therefore a higher content of QM for massive NS.}
 \label{FIG:PSR}
\end{figure}
\begin{figure}[h]
 \includegraphics[scale=0.45,angle=-90]{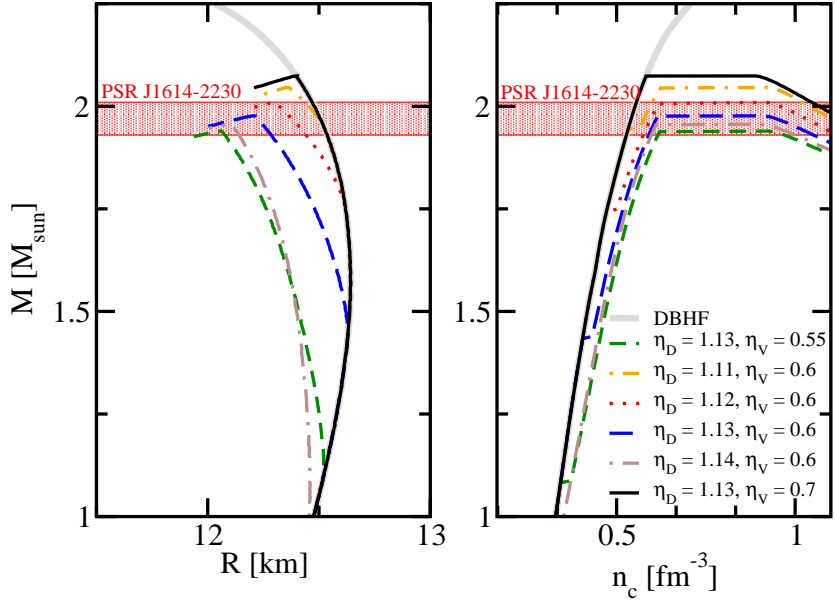}
 \caption{Mass-radius and mass-central density relation for a selection 
of EoS parameterizations with mostly significant QM core 
(as found in the upper right corner of Fig. \ref{FIG:MonsterPlot}).}
 \label{FIG:GoodHybrids}
\end{figure}

It is interesting to observe,
that there is a band (red) of EoS parameters
in the $\eta_D$-$\eta_V$ plane
with resulting maximum NS masses of $1.97\pm0.04$ M$_\odot$.
While the red band in Fig.~\ref{FIG:MonsterPlot}
denotes configurations with {\it maximum} masses
corresponding to PSR J1614, it is worthwhile
to ask for the actual quark content of these
configurations.
For the purpose of this article we will make
a qualitative statement, only, and postpone
quantitative analyses to a later publication.
The curves labeled $M_{\rm onset}$ in Fig.~\ref{FIG:MonsterPlot}
indicate from which NS mass on the EoS results in 
hybrid NS solutions. 
Hence it is favorable  to have a small
value of $M_{\rm onset}$ in order to obtain
large QM cores.
Therefore, the largest QM content for a high mass NS
has to be expected in the upper right corner of Fig.~\ref{FIG:MonsterPlot}.
This is illustrated in Fig.~\ref{FIG:PSR}.
We point out, that the red band describes NS configurations
where the maximum mass corresponds to the reported mass of PSR J1614, only.
Of course, more massive NS are possible.
As we are interested in massive NS with large quark cores
Fig.~\ref{FIG:GoodHybrids} shows NS configurations from the upper right corner 
of Fig.~\ref{FIG:MonsterPlot}.
Large quark cores in our EoS parameterization are in general favored for 
values of $\eta_V=0.5\dots0.6$.
This is in vicinity of the value $\eta_V^{F}=0.5$ as obtained after Fierz 
transformation of the one-gluon exchange interaction (details in 
Ref.~\cite{hep-ph/0402234}).
It is noteworthy, that comparable small changes of the diquark coupling 
constant strongly affect the transition density.
As an example, Fig.~\ref{FIG:GoodHybrids} illustrates, that with constant 
$\eta_V=0.6$ a change of a few percent in $\eta_D$ ($=1.11\dots 1.14$) makes 
the difference whether a rather typical NS with a mass of, e.g., 
M$=1.4$ M$_\odot$ has QM content or not.

\subsection{Implications for HIC}

\begin{figure}[h]
 \includegraphics[scale=0.45,angle=-90]{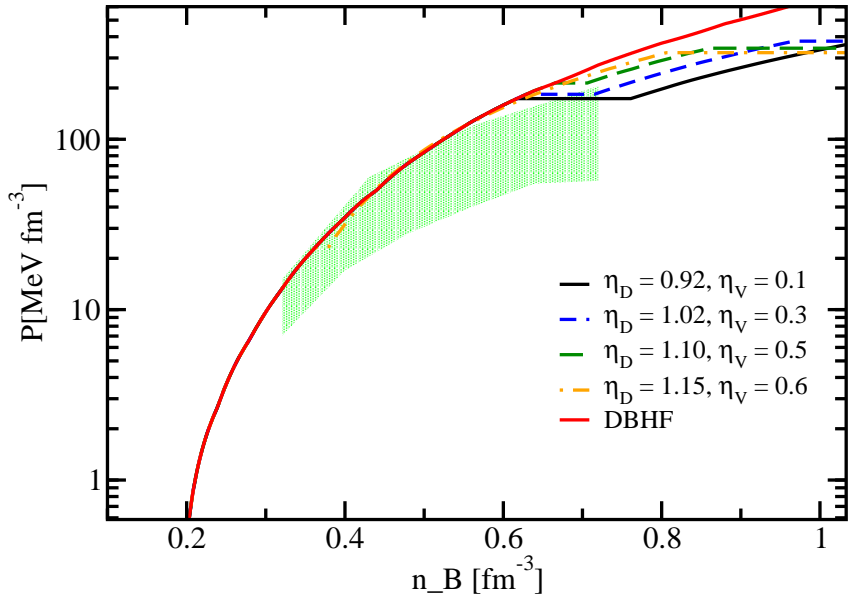}
 \caption{Flow constraint of symmetric equation of state for chosen parameter 
sets with maximum masses equal to that of PSR J1614-2230.}
 \label{FIG:flowPRS}
\end{figure}
\begin{figure}[h]
 \includegraphics[scale=0.45,angle=-90]{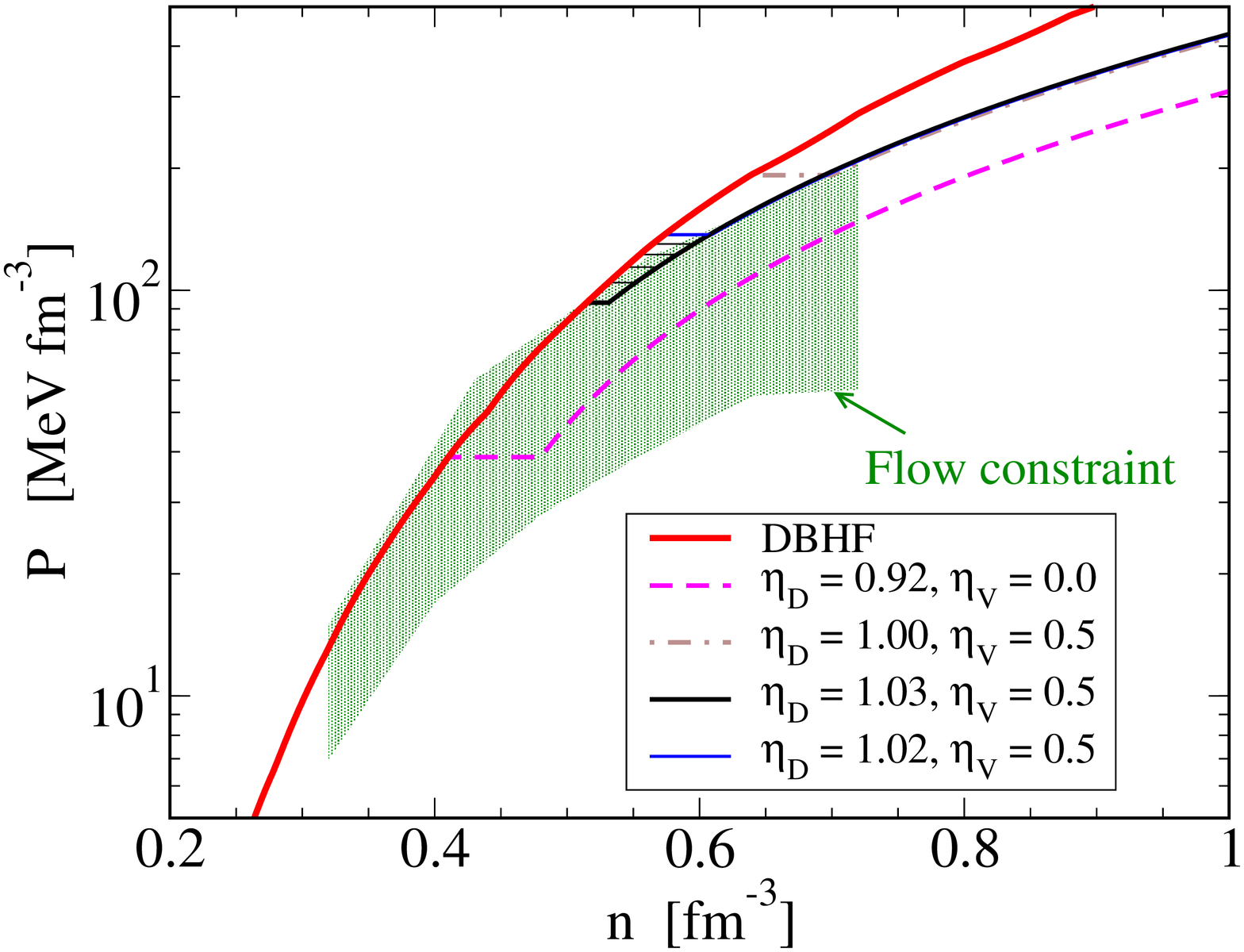}
 \caption{In previous work \cite{Klahn:2006iw} better agreement with the flow 
constraint has been achieved. Reason is a different parameterization which in 
particular results in a smaller dynamical quark mass in vacuum.}
\label{FIG:Flow2006}
\end{figure}
In isospin-symmetric matter as found in HIC our main statements
regarding the influence of $\eta_V$ and $\eta_D$ on the stiffness
of the EoS and the transition density do not change.
Increasing $\eta_V$ at constant $\eta_D$ increases the stiffness and the 
transition density, increasing $\eta_D$ at constant $\eta_V$ reduces the 
stiffness and lowers the transition density.
The transition densities in symmetric matter are plotted in 
Fig.~\ref{FIG:MonsterPlot}, labeled as n$_{\rm onset}^{\rm sym}$.
It is remarkable that along the red band (configurations
with maximum masses corresponding to the mass of PSR J1614)
the transition density in symmetric matter has an almost constant value
of n$_{\rm onset}^{\rm sym}\approx4$n$_S$.
This is a distinct difference to what we learned for the electrically neutral 
and $\beta$-equilibrated EoS for NS matter, where the transition density
decreased from the lower left to the upper right within the red band.
Since the transition density in symmetric matter is lower only below the red 
band, a region where the maximum NS masses are not sufficiently large
this model predicts n$_{\rm onset}^{\rm sym}\approx4$n$_S$ as
a lower limit on the critical density for the phase transition
in symmetric matter.
It is necessary to point out, that our previous study \cite{Klahn:2006iw} gave 
a different result and predicted a lower transition density.
In order to illustrate this, we plot hybrid EoS parameterizations in symmetric 
matter corresponding to NS EoS with maximum masses of $1.97\pm0.04$ M$_\odot$ 
(along the red band in Fig.~\ref{FIG:MonsterPlot}) in 
Fig.~\ref{FIG:flowPRS} and our previous result in Fig.~\ref{FIG:Flow2006}.
While Fig.~\ref{FIG:flowPRS} suggests, that the phase transition
occurs at densities too high to prevent the nuclear DBHF EoS from
violating the flow constraint, our old results in Fig.~\ref{FIG:Flow2006} 
perform significantly better.
This observation gives our study a surprising twist which we will
investigate in future work.
The reason for this different behavior is not hard to comprehend.
Both EoS start from a different parameterization already on the level
of the scalar coupling constant.
The larger dynamical quark mass
resulting from the parameterization applied for
the present study (see \cite{Grigorian:2006qe} for the actual values) 
causes a general shift of the QM onset to larger chemical potentials. 
Therefore, the phase transition from nuclear to QM simply follows this trend.
From our perspective, this is the first time, that
high density observables as elliptic flow and maximum NS masses
can actually be applied to distinguish models
with different parameterizations in the scalar channel.
Of course, the usual caveats have to be made, lead by the most serious one:
In order to draw final conclusions first it has to be
made clear beyond any doubt, that NS actually have a QM core.
At the current stage, this statement is neither proven nor disproven.
If observational evidence in favor of the presence of QM in NS
should ever arise, many more statements can be made,
not only about the transition density in symmetric matter.
Large NS masses require rather stiff QM EoS.
As a consequence, the transition region in the density domain is not very 
large. 
This is illustrated Table~\ref{TAB:transDjump} by a few examples 
for parameters within the red band in Fig.~\ref{FIG:MonsterPlot}) 
by calculating the difference $\Delta$n
between QM and NM density at the phase transition.
Since the quark content increases with $\eta_V$
a clear signature for QM in NS would imply at least 
$\Delta$n$<$n$_S\approx$0.16 fm$^{-3}$.
\begin{center}
\begin{table}[h]
\begin{tabular}{|c|lllll|}
 \hline
 $\eta_V$ 			& 0.1 	& 0.2 	& 0.3 	& 0.5 	& 0.6 \\
 $\eta_D$ 			& 0.92 	& 0.98 	& 1.02 	& 1.10 	& 1.15 \\\hline
 $\Delta$n [fm$^{-3}$] 	& 0.144 	& 0.107	& 0.086	& 0.041	&  0.017 \\\hline
\end{tabular}
 \caption{Density difference between QM and NM at the phase transition 
for a few parameterizations with M$_{\rm max}\in (1.97\pm0.04)$M$_\odot$.}
 \label{TAB:transDjump}
\end{table}
\end{center}

%
%

%
%
\section{Conclusions}
\label{SEC:CONCL}
We performed a systematic model analysis of
an NJL-type EoS where we investigated the
dependence of NS properties on the 
diquark and vector coupling constants.
It is possible to describe NS configurations
with a significant amount of QM and maximum NS masses
which are in agreement with the reported high mass
of PSR J1614-2230.
The amount of QM in a NS will increase
with increasing vector coupling 
if the diquark coupling is adjusted accordingly.
Therefore, both channels are important for
the understanding of NS phenomenology.
Due to a different choice of the scalar coupling
and the resulting larger dynamical quark mass
the present model parameterization does not resolve the conflict of the nuclear
DBHF EoS with the flow constraint.
This result is of great importance since it implies
that without a sound understanding of the
chiral phase transition in medium otherwise identical
models can come to quantitatively significantly
different results.
In particular this raises doubts about the
predictive power of any EoS which does not
account for the mechanism of chiral symmetry breaking.
Of course, we do not intend to imply, that NJL-type models
will provide the ultimate tool to describe
the QCD-phase transition qualitatively and quantitatively
correct.
There are many open questions, which will need to be
addressed in the future, most importantly the mechanism
of confinement/deconfinement in medium.
Significant improvement of any existing model
is required in order to understand this phenomenon
and to finally get rid of the necessity to
construct thermodynamically motivated Maxwell-
and Gibbs-phase transitions, which are not suited to provide
deeper insights into the microphysical mechanisms governing 
QCD phase transformations.
First steps in this direction have been performed within
a generalized NJL model \cite{nucl-th/0602014}.
Further progress was made in a microscopic description
of the baryon dissociation in dense matter, based again on 
a NJL model approach \cite{arXiv:1008.4029} where in particular the 
interplay between chiral symmetry restoration and diquark condensation 
transitions at high densities for the spectral
function of nucleons has been investigated.
Further steps will be taken
towards the development of an EoS which describes
the Mott dissociation of baryons 
into their quark constituents in a consistent way
which accounts for  nucleonic bound and scattering states simultaneously.
An  EoS of this quality has been developed
within a  generalized Beth-Uhlenbeck approach
for strongly interacting matter
in order to describe the Mott dissociation of deuterons in nuclear matter \cite{Schmidt:1990}
(see also recent work on cluster formation in low-density nuclear matter
\cite{arXiv:0908.2344,Hempel:2011kh})  
and the Mott effect for mesons in quark matter \cite{386637,384452}.
We are convinced that studies of the Mott mechanism for the dissociation of 
baryons in dense matter which generalize early nonrelativistic approaches 
\cite{Blaschke:1984yj,Ropke:1986qs}
to a field theoretical formulation
will illuminate the role of this mechanism 
in its interplay with chiral symmetry restoration and 
diquark condensation for the EoS of dense matter and its applications 
in HIC and in astrophysical environments.

%
%

\subsection*{Acknowledgement}
The authors acknowledge valuable discussions and collaboration, in particular 
with J.~Berdermann, M.~Buballa, H.~Grigorian, G.~Poghosyan, C.~D.~Roberts, 
G.~R\"opke, F.~Sandin, S.~Typel, D.~N.~Voskresensky, F.~Weber, H.~Wolter and 
D.~Zablocki.
This work has been supported in part by ``CompStar'', a Research Networking 
Programme of the European Science Foundation and by the Polish Ministry for
Science and Higher Education under grant No. NN 202 2318 37.
The work of D.B. has been supported by the Russian Fund for Fundamental 
Investigations under grant No. 11-02-01538-a.
R.{\L}. received support from the Bogoliubov-Infeld programme
for visits at the JINR Dubna where part of this work was done.

\end{document}